%% file: main.tex
\documentclass{sig-alternate}

\pdfoutput=1
\usepackage[utf8]{inputenc}
\usepackage[T1]{fontenc}
\usepackage{color}
\usepackage{amsmath}
\usepackage{enumerate}
\usepackage{url}
\usepackage{multirow}
\usepackage{rotating}
\usepackage{threeparttable}
\usepackage{pgfplots}
\usetikzlibrary{patterns}
\usetikzlibrary{calc}

\providecommand{\e}[1]{\ensuremath{\times 10^{#1}}} 

\newtheorem{definition}{DEFINITION}

\begin{document}

\title{Predicting the Temporal Dynamics of Information Diffusion in Social Networks\footnotemark[1]}

\numberofauthors{3} 
\author{
\alignauthor
Adrien Guille\\
       \affaddr{ERIC Laboratory, Lyon 2 University}\\
	   \affaddr{5 av. Pierre Mendes France, 69676 Bron, France}\\	   				
       \email{Adrien.Guille@univ-lyon2.fr}
\alignauthor
Hakim Hacid\\
       \affaddr{Bell Labs France}\\
	   \affaddr{Route de Nozay, 91620 Nozay, France}\\	   				
       \email{Hakim.Hacid@alcatel-lucent.com}      
\alignauthor
C\'ecile Favre\\
	   \affaddr{ERIC Laboratory, Lyon 2 University}\\
	   \affaddr{5 av. Pierre Mendes France, 69676 Bron, France}\\	   				
       \email{Cecile.Favre@univ-lyon2.fr}      
}    

\maketitle

\begin{abstract}
\input{abstract}
\end{abstract}

\section{Introduction}
\footnotetext[1]{This paper is an updated and extended version of \cite{guille2012}.}

\input{introduction}

\section{Related work}
\label{sec:related_work}
\input{related_work}

\subsection{Diffusion in complex systems}

\textbf{Graph based approaches.}
\input{graphical_models}

\textbf{Non-graph based approaches.}
\input{nongraphical_models}

\subsection{Information diffusion in OSNs}
\input{predictive_osn}

\section{Diffusion observation and representation}
\label{sec:data}
\input{data}

\subsection{Topic extraction and diffusion observation}
\input{dataset_extraction}

\section{Proposed method}

\label{sec:method}
\input{proposed_approach}

\subsection{Time-Based Asynchronous Independent Cascades}
\input{t-basic}

\subsection{Features space}
\input{features}

\subsection{Diffusion function parameters estimation}
\input{diffusion_function}

\subsection{Time-delay estimation}
\input{time-delay}

\subsection{Generalization of the approach}
\input{prediction_engine}

\input{generalization}

\section{Experiments}
\label{sec:experiments}

\input{experiments}

\subsection{Qualitative results}
\input{qualitative_results}

\subsection{Quantitative results}
\input{quantitative_results}

\section{Conclusion}
\label{sec:conclusion}
\input{conclusion}


\end{document}

%% file: abstract.tex
Online social networks play a major role in the spread of information at very large scale and it becomes essential to provide means to analyse this phenomenon. In this paper we address the issue of predicting the temporal dynamics of the information diffusion process. We develop a graph-based approach built on the assumption that \textit{the macroscopic dynamics of the spreading process are explained by the topology of the network and the interactions that occur through it, between pairs of users, on the basis of properties at the microscopic level.} We introduce a generic model, called \textit{T-BaSIC},  and describe how to estimate its parameters from users behaviours using machine learning techniques. Contrary to classical approaches where the parameters are fixed in advance, \textit{T-BaSIC}'s parameters are functions depending of time, which permit to better approximate and adapt to the diffusion phenomenon observed in online social networks. Our proposal has been validated on real Twitter datasets. Experiments show that our approach is able to capture the particular patterns of diffusion depending of the studied sub-networks of users and topics. The results corroborate the ``two-step'' theory (1955) that states that information flows from media to a few ``opinion leaders'' who then transfer it to the mass population via social networks and show that it applies in the online context. This work also highlights interesting recommendations for future investigations.

%% file: introduction.tex
The Web 2.0 -- through the concepts of ``participatory'' and ``social'' web -- allows hundreds of millions of Internet users worldwide to produce and consume content. Thus, the Web provides access to a very vast source of information on an unprecedented scale. Online social networks play a major role in the diffusion of this information and have proven to be very powerful in many situations, such as the 2010 Arab Spring \cite{arabspring} or the 2008 U.S. presidential elections \cite{hughes2009twitter}. They permit people to spread ideas, to organize groups and actions in a new way and we can consequently consider that they add a whole new layer to the human social life. In consideration of the impact of online social networks on the society, understanding the mechanics and dynamics of these networks is a critical research objective.
Since communications occurring online are recorded, very large amounts of data are available for researchers who can exploit them to develop predictive models for information diffusion in online social networks. This proves to be a challenging task, due to (i) the particular laws that govern these networks, (ii) the wide diversity in users profiles and, obviously, (iii) the large scale of these structures. 

``Information diffusion'' is a generic concept that refers to all processes of propagation in a system, regardless of the nature of the object in motion. The diffusion of innovation over a network is one of the original reasons for studying networks and the spread of disease among a population has been studied for centuries. The models developed in the context of social networks assume that people are influenced by actions taken by their surrounding in the network, in other words, they model processes of ``information cascades'' \cite{information_cascade}. That is why the path followed by an information in the network is often referred to as the ``spreading cascade''. In the case of online social networks, one can be interested by the spread of particular objects like hashtags on Twitter, URLs, or even broader concepts like topics for instance.\\ 

\noindent\textbf{PROBLEM DEFINITION}: \textit{Having a set of users in a social network (with explicit or inferred connexions), communicating through a messaging system in a closed environment, and a piece of information, how to predict the degree of adoption of such information in the provided social network for a given period of time, i.e. the temporal patterns of the dynamics?\\}

A closed environment here means that only internal constraints are considered. For example, we don't take into account the possibility that information may come from external sources like news sites.
Modeling information diffusion first requires to define the set of actors that can potentially be involved. In the context of online social networks, an actor is referred to as a node. The simplest way to describe the spreading process is to consider that a node can be whether activated or not (i.e. informed or not), and then, the propagation process can be viewed as a sequence of activation of nodes. A diffusion process occurring inside a network is characterized by two aspects: (i) its topology and (ii) its temporal dynamic.

Understanding, capturing, and being able to predict such phenomenon can be helpful for several areas such as marketing, security, and Web search. These use-cases fall all under either of these two well defined problems: (i) influence maximization \cite{LT}, e.g. maximizing spread of information, and (ii) influence minimization, e.g. minimizing spread of misinformation \cite{misinformation, blocking}. 
Most of existing predictive models focus on the topology of the process and are based on uni-dimensional feature spaces. They intend to predict properties like the depth of the spreading cascade or the total size of the reached population and vastly ignore the temporal dimension. In this paper, we consider the issue of predicting the temporal dynamics of the diffusion process -- more specifically the spread of topics -- in online social networks. Our initial assumption is that \textit{the macroscopic dynamics (i.e. observed overall the network) of the spreading process are explained by the topology of the network and the interactions that occur through it, between pairs of users, on the basis of properties at the microscopic level (localized in the network)}. The contributions of this paper are the following: 

\begin{enumerate}
\item An analytical discussion about how to detect spreading topics and the features that may explain the diffusion process. This step has been performed using a dataset crawled from Twitter. It enabled us to understand the overall process of information diffusion in a real social network.
\item A new model for information diffusion modeling in online social networks and a set of features used to estimate its parameters. This new model, \textit{T-BAsIC}, permits a deeper and realistic integration of time in the prediction process. The features are based on users behaviour and belong to three dimensions (social, topic, and time).
\item An experimental evaluation that aims to assess the efficiency of our approach and the validity of the underlying assumption (i.e. the macroscopic diffusion process is explained by the sum of microscopic interactions that occur because of local properties).
\end{enumerate}

The rest of this paper is organized as follows. Section~\ref{sec:related_work} reviews various categories of related work and discusses their relation to ours. In Section~\ref{sec:data} we present the data and the analysis we performed. Then in Section \ref{sec:method} we describes the proposed model in details. In Section \ref{sec:experiments}, a set of experiments is described to evaluate the efficiency of our modeling and the validity of the underlying assumption. We conclude and provide some future work in Section \ref{sec:conclusion}.

%% file: related_work.tex
In this section we review two categories of related work: (i) general modeling of spreading processes in complex systems, classified into graphical and non-graphical approaches, and (ii) recent predictive models of information diffusion in online social networks (OSNs).

%% file: graphical_models.tex
Classical graph based approaches assume the existence of a graph structure  and focus on the topology of the process. They follow either \textit{Independent Cascades} (IC) \cite{IC} or \textit{Linear Threshold} (LT) \cite{LT} model. They are based on a directed graph where each node can be activated (i.e. informed) or not, with a monotonicity assumption, meaning that activated nodes cannot deactivate. The IC model requires a diffusion probability to be associated to each edge whereas LT requires an influence degree to be defined on each edge and an influence threshold for each node. For both models, the diffusion process proceeds iteratively in a synchronous way along a discrete time-axis, starting from a set of initially activated nodes. In the case of IC, for each iteration, the newly activated nodes try once to activate their neighbours with the probability defined on the edge joining them. In the case of LT, at each iteration, the inactive nodes are activated by their activated neighbours if the sum of influence degrees exceeds their own influence threshold. Successful activations are effective at the next iteration. In both cases, the process ends when no new transmission is possible, i.e. no neighbouring node can be contacted. These two mechanisms reflect two different points of view: IC is sender-centric while LT is receiver-centric. Both models have the inconvenience to proceed in a synchronous way along a discrete time-axis, which doesn't suit what is observed in real social networks. In order to make these models more adapted to this particular context, Saito et al. recently proposed asynchronous extensions of these models, namely \textit{AsIC} and \textit{AsLT} \cite{asic}, that use a continuous time-axis and require a time-delay parameter on each edge of the graph.

%% file: nongraphical_models.tex
Classical non-graph based approaches don't assume the existence of a graph structure and have been mainly developed to model epidemiological processes. They classify nodes into several classes (i.e. states) and focus on the evolution of the proportions of nodes in each class. The two most common models are \textit{SIR} and \textit{SIS} \cite{sir_sis,newman}, where $S$ stands for ``susceptible'', $I$ for ``infected'' (i.e. informed) and $R$ for recovered (i.e. refractory). In both cases, nodes in the $S$ class switch to the $I$ class with a fixed probability $\beta$. Then, in the case of $SIS$, nodes in the $I$ class switch to the $S$ class with a fixed probability $\gamma$, whereas in the case of $SIR$ they permanently switch to the $R$ class. The percentage of nodes in each class is given by simple differential equations. Both models assume that every node has the same probability to be connected to another and thus connections inside the population are made at random. But the topology of the nodes relations is very important in social networks and thus the assumptions made by these model are unrealistic.

%% file: predictive_osn.tex
Various studies in the context of social networks have been conducted with the aim of predicting properties of the information spreading process. Most of them focus on topological properties. For instance, Bakshy et al.~\cite{everyone} proposed a graphical approach that aims to predict the size of the cascade generated by the diffusion of a URL in Twitter graph of followers, starting with a single initial user. This sender-centric model relies on a regression tree and some simple social attributes and the past influence of the initial user only. The influence of the initial user is approximated by counting the number of cascades (implicit cascades inferred from the follower graph) in which he was involved in the past. Galuba et al. \cite{galuba} also studied the diffusion of URLs in Twitter, but from a receiver-centric point of view, and proposed to use the LT model to predict which users will adopt which URL. Yang and Counts \cite{microsoft} used survival analysis to examine the impact of attributes from both users and content to predict the size of the cascades generated by the spread of topics in Twitter. In order to do so, they exclusively focused on targeted tweets so they can directly identify the explicit cascade of diffusion. They found that both user and content attributes were relevant predictors of the diffusion efficiency.

To the best of our knowledge, the Linear Influence Model developed by Yang and Leskovec~\cite{lim} is the only real predictive model for the temporal dynamics that has already been proposed. They studied the diffusion of hashtags in Twitter and proposed a model based on the assumption that the influence of a node depends on how many other nodes it influenced in the past. However, there is a substantial difference with our work because this approach is non-graph based and doesn't study nodes attributes. Therefore, this approach doesn't take advantage of any knowledge about the topology of the network. Moreover, in their modeling, a node corresponds to the aggregation of 100 Twitter users, which doesn't permit to study the diffusion at a ``user to user'' level. 

Given this state-of-the art, we propose a generic \textit{graph-based} method, so the topology of the network is exploited, to model information diffusion. We also detail how to apply it on Twitter to predict the \textit{temporal dynamics} of the spread of \textit{topics} among users. 
This is a particularly interesting contribution of our work since all existing approaches have been mainly focusing on the prediction of depth of the cascades and/or the final volume of the propagation.

%% file: data.tex
In this section, we discuss some observations we have performed in order to understand the diffusion process in social networks and extract some underlying facts to represent such phenomenon. 
For availability reasons mainly, we -- like the majority of studies that address information diffusion modeling in social networks which have used non-synthetic data -- build the observation part of this paper on data coming from Twitter. This allows us to easily position and compare our approach with related work.
Twitter is a micro-blogging service that allows its users to publish public directed or undirected short messages (140 characters at most) and to follow other users that interest them. Send a directed message is achieved by mentioning the targeted users directly in the content with the convention ``@username''. Both directed and undirected messages are automatically forwarded to the followers but directed messages aim more particularly at one or more specific users. Overall, Twitter forms an online social network where the information flows from place to place, in two different ways: (i) it flows in a passive manner via the following ties and also (ii) actively because of users that directly send information to others via the mentioning practice. 
Mathematically, this network is represented as a directed multi-graph comprised of two sets of edges: (i) the set of following edges, which constitutes the \textit{passive} part of the network, commonly called ``follower graph''; (ii) the set of mentioning edges, that represents the \textit{active} part of the social network.

\begin{definition}[Active/Passive Directed Edge] 
An active edge results from an explicit communication (i.e. message passing) between two nodes in the network. It translates the existence of an active transmission of information between two nodes. A passive edge simply means that a node is exposed to the content produced by another.
\end{definition}

We base our study on a 467 million Twitter posts dataset from 20 million users covering a 7 months period from June 1, 2009 to December 31, 2009 gathered by Yang and Leskovec \cite{twitter7}. Each tweet contains its author, its content, and the time at which it has been posted. In addition to that, we know the sets of followers and followees for each user thanks to the capture of Twitter graph of followers (1.47 billion directed edges) made by Kwak et al.~\cite{kwak} at the same time (i.e. passive edges). Finally, we complete the network by extracting the active edges based on the mentions contained in the tweets. This data meets the common criteria for dataset validation in social network analysis: (i) large-scale, (ii) completeness, and (iii) realism.
A key task in the diffusion model we are proposing is the identification and detection of topics, a process explained in what follows. 

%% file: dataset_extraction.tex
We intend to predict the spread of information among Twitter users. In contrast to studies that have investigated the diffusion of simple objects, such as URL or hashtags \cite{lim,everyone,galuba}, we focus on topics as the main object to follow. This allows us to have a global view of all the interactions regarding a specific information. This also prevents from several annoyances, like the side effects that potentially exist between distinct URLs that point to similar resources for instance. By cons, it is not so easy to detect spreading topics.
For the purpose of our study, we define a topic as follows: 

\begin{definition}[Topic] 
A topic is a minimal set of co-occurring terms (i.e. keywords) that a related tweet should contain and which spans over a given period of time.
\end{definition}

We are interested by recurrent terms that experience a peak in their usage during a significant period of time. It means that we are not interested by non-recurrent terms that are not observed before and after the period during which they are popular. To find topics that fit this definition, we use the method described hereafter to find relevant terms and then manually investigate further to precisely define interesting set of terms. 
We select all the tweets published during the period of time we want to study and perform a discretization. To do so, the data is transformed into an ordered collection of documents, where each document is the aggregation of 4 hours of tweets\footnote{This is done using the \textit{Lucene} (\url{http://lucene.apache.org/core/} library to index their content).}. Then we compute the vector of the number of occurrences of a term in each document, noted $O_{term}$. We define the interestingness of a term as the score computed by Equation~\ref{eq:topics}. Note that highest values are obtained by terms that maximize the ratio  $\max(O_{term})/\text{avg}(O_{term})$ and minimize the ratio $\min (O_{term})/\text{avg}(O_{term})$.
Finally, we rank the terms according to their score in order to identify which topics to focus on.

\begin{equation}
\label{eq:topics}
score(term) = \frac{\text{avg}(O_{term})^2+\min (O_{term})\times \max (O_{term})}{\min (O_{term})\times \text{avg}(O_{term})}
\end{equation}
	
\begin{table}
\center
\begin{tabular}{|c|c|}
\hline \textbf{term} & \textbf{score} \\ 
\hline christmas & 24.56 \\
\hline snow & 22.88 \\
\hline iphone & 15.19 \\
\hline google & 15.05 \\ 
\hline ... & ... \\
\hline twitpic.com & 6.64 \\
\hline twitter & 6.34 \\
\hline bit.ly & 5.14 \\
\hline lol & 5.12 \\
\hline 
\end{tabular} 
\caption{Highest and lowest ranked terms according to the interestingness score during the month of December 2009.}
\label{tab:keywords}
\end{table}

See Table \ref{tab:keywords} for the 4 best and worst ranked terms in 2009. Unsurprisingly, ``christmas'' is the top ranked term, because it is a sustained discussion topic throughout the month and suddenly becomes an extremely popular term right before and after Christmas on December 25$^{\text{th}}$. Therefore, this is a particular case where the peak of activity is linked to an annual event and doesn't result from the spread of an interesting information between Twitter users. Let's consider now the example of the term ``iphone''. 
We observe a peak of usage starting around December 8$^{\text{th}}$ on Figure \ref{fig:terms}(b). By searching through specialized websites, we find that a rumour about the \textit{``release''} date of the new version of the smartphone surfaced on this day and then spread through social networks. This is confirmed by a sharp increase in the frequency of co-occurrence of these two terms (\textit{``iphone'', ``release''}) in tweets published from December 8$^{\text{th}}$ to 15$^{\text{th}}$. Therefore, the set \{``iphone'',``release''\} defines an interesting spreading topic. 
It is the same for the term ``google'', that experiences a peak of activity in December because of the spread of a rumor about the buyout of a company whose technology could contribute to Google Wave (thus we define the topic \{``google'',``buy''\}). On the contrary, ``twitpic.com'' has a relatively steady volume (because it appears each time a user posts a picture with her tweet) and is therefore bad ranked.

\pgfplotsset{width=7.5cm,height=3.5cm}
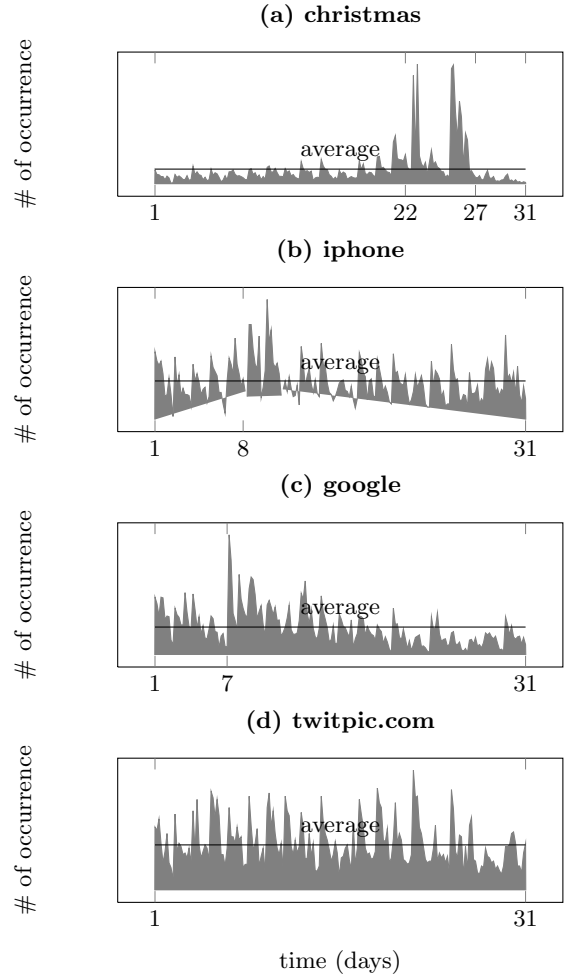
\begin{figure}[ht]
\begin{center}
\begin{tikzpicture}
\begin{axis}
[title = \textbf{(a) christmas},name=christmas,
xtick \empty,
ytick \empty,
extra x ticks={0,125,160,185},
extra x tick labels={1,22,27,31},
ylabel= \# of occurrence,
bar width=0.025pt]
\addplot [black,fill=black!90!black,opacity=0.5] table 
[x=time, y=christmas] {christmas.dat};
\addplot[black,sharp plot]
coordinates {(0,67) (185,67)}
node[above] at (axis cs:92.5,67) {average};
\end{axis}
\end{tikzpicture}

\begin{tikzpicture}
\begin{axis}
[title = \textbf{(b) iphone},name=iphone,
xtick \empty,
ytick \empty,
ylabel= \# of occurrence,
extra x ticks={0,44,185},
extra x tick labels={1,8,31},
bar width=0.025pt]
\addplot [black,fill=black!90!black,opacity=0.5] table 
[x=time, y=iphone] {iphone.dat};
\addplot[black,sharp plot]
coordinates {(0,37) (185,37)}
node[above] at (axis cs:92.5,36) {average};
\end{axis}
\end{tikzpicture}

\begin{tikzpicture}
\begin{axis}
[title = \textbf{(c) google},name=google,
xtick \empty,
extra x ticks={0,36,185},
extra x tick labels={1,7,31},
ytick \empty,
ylabel= \# of occurrence,
bar width=0.025pt]
\addplot [black,fill=black!90!black,opacity=0.5] table 
[x=time, y=google] {google.dat};
\addplot[black,sharp plot]
coordinates {(0,43) (185,43)}
node[above] at (axis cs:92.5,43) {average};
\end{axis}
\end{tikzpicture}

\begin{tikzpicture}
\begin{axis}
[title = \textbf{(d) twitpic.com},
xtick \empty,
extra x ticks={0,185},
extra x tick labels={1,31},
ylabel= \# of occurrence,
xlabel=time (days),
ytick \empty,
bar width=0.025pt]
\addplot [black,fill=black!90!black,opacity=0.5] table 
[x=time, y=twitpic] {twitpic.dat};
\addplot[black,sharp plot]
coordinates {(0,28) (185,28)}
node[above] at (axis cs:92.5,28) {average};
\end{axis}
\end{tikzpicture}
\caption{Evolution of the volume of particular terms during December 2009, namely, from top to bottom, ``christmas'', ``iphone'', ``google'' and ``twitpic.com''. }
\label{fig:terms}
\end{center}
\end{figure}

Through this analysis, at least two dimensions that are needed to capture the diffusion process have emerged: (i) the \textit{topical} dimension since we observed that the various topics had different behaviours in terms of volumes for instance, meaning that users are not interested in all topics but generally in a subset of those topics; (ii) the \textit{temporal} dimension, because we observed a common cyclic pattern to all topics that is due to, e.g. the switch from day to night, working hours, and the total time for the spread of the information.

\subsection{Representation of the propagation process}

We exploit the data related to selected topics to observe aspects of the diffusion process in Twitter. Thus, we build the structure that transcribes ``who influenced whom'' for each topic and we also capture the temporal dynamics of the diffusion. This process is built on two concepts: \textit{activation sequence} and \textit{spreading cascade}. 

\begin{definition}[Activation Sequence] 
An activation sequence is an ordered set of nodes capturing the order in which the nodes of the network adopted the topic, i.e. got informed.
\end{definition}

\begin{definition}[Spreading Cascade] 
A spreading cascade is a tree having as a root the first node of the activation sequence. The tree captures the influence between nodes (i.e. who transmitted the information to whom) and unfolds in the same order as the activation sequence.
\end{definition}

\begin{figure}[]
\begin{center}
\includegraphics[width=0.60\linewidth ]{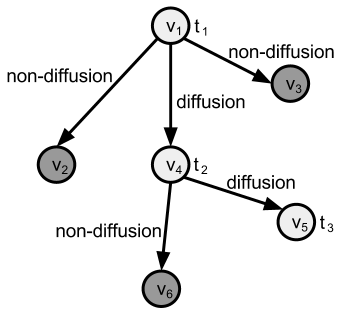}
\caption{Example of a spreading cascade. Nodes colored in light gray represent users that have tweeted about the topic of interest at time $t$, with $t_1<t_2<t_3$. Users represented by nodes coloured in dark grey didn't.}
\label{fig:cascades}
\end{center}
\end{figure}

Having a topic and its minimal set of keywords, we can easily detect all the related messages and generate the time-series of the volume of tweets induced by the diffusion of the topic (i.e. its dynamics). We also determine the sequence of nodes activation in the network. Then we aim to solve the problem of reconstructing the graph of diffusion (i.e. the spreading cascade) by connecting the activated nodes between them. We base the construction of this graph on the topology of the passive part of the network. In other words, we model the spreading process over the following links. It means that for each activated node, we want to infer which other previously activated node among its followees had influenced her. As it is discussed in \cite{bakshy}, in the case where several followees are activated, there are basically three ways to assign influence: (i) assign it to the followee that adopted the topic first, (ii) assign it to the last followee to react, or (iii) assign it to all the followees. In this study we assume that individuals are influenced by the followee that adopted the topic most recently (i.e. the second method, referred to as ``Last Influence''). Let us illustrate how we build the spreading cascade of a topic with the following example. Let's say we have a social network of 6 users, where $v_2$, $v_3$, $v_4$ and $v_5$ follow $v_1$; $v_5$ and $v_6$ follow $u_4$. Nodes $v_1$, $v_4$ and $v_5$ are activated in this order. Therefore, based on the ``Last Influence'' principle, we say that instances of diffusion have occurred between $v_1$ and $v_4$, and $v_4$ and $v_5$, whereas there are instances of non-diffusion between $v_1$ and $v_4$, $v_4$ and $v_6$, etc. Finally we can build the spreading cascade shown on the Figure \ref{fig:cascades}, where each edge is directed and labeled with either ``diffusion'' or ``non-diffusion''.

With the methods we described, we are able to (i) detect interesting spreading topics and (ii) capture their diffusion process. Thus we can construct datasets for various topics, consisting of instances belonging to the binary class \{diffusion, non-diffusion\} and described by a pair of users and a timestamp. Moreover, a third dimension is explicitly highlighted thanks to this representation: the \textit{social} dimension since the information flows due to influence between members of the social network. As a result, the three dimensions (social, topic, time) are the foundations of the model we are proposing and which we describe in the next section. 

%% file: proposed_approach.tex
\input{notations}

In this section we introduce the method we propose to predict the diffusion phenomenon observed in online social networks. To start, we formally define the \textit{Time-Based Asynchronous Independent Cascades (T-BAsIC) Model} underlying our approach. Then, we present the list of features computed for each member of the network and finally we describe how they are used to estimate the model parameters. Table \ref{tab:notations} summarizes the notations used in this section.

%% file: notations.tex
\begin{table*}[ht]
\centering
\begin{tabular}{|c|l|}
\hline \textbf{Notation} & \textbf{Description} \\ 
\hline $V$ & the set of all vertices (i.e. users) in the social network \\ 
\hline $v_x \in V$ & a particular node in the social network\\ 
\hline $S$ & a subset of vertices $S \subset V$ \\ 
\hline $E$ & the set of all edges in the social network \\ 
\hline $M$ & all the messages (i.e. tweets) of the environment \\ 
\hline $M_{v}$ & the set of messages published by a user $v \in V$ \\ 
\hline \small ${\cal{M}}_v$ & the set of users who \textit{are mentioned} in the messages of a user $u \in U$ \\ 
\hline $\overline{{\cal{M}}_v}$ & the set of users who \textit{mentioned}  the user $v\in V$ in their messages\\
\hline $t{\cal{M}}^v$ & all the messages which have mentioned a user $v \in V$ \\ 
\hline $K$ & the set of all keywords used in the messages published inside the network \\ 
\hline $k_i \in K$ & a specific keyword contained in the messages published inside the network \\ 
\hline $K_v \subset K$ & the set of keywords included in the messages published by a user $v \in V$ \\ 
\hline $\cal{C} = \{ $$c_{1}, c_{2}, ..., c_{g}$$\}$ & the set of all the topics. $c_i$ is a particular topic. \\ 
\hline $c_i = \{k_1,k_2, ..., k_p\}$ & the vector of keywords $k_j \in K$ describing a topic $c_i$ \\ 

\hline 
\end{tabular} 
\caption{Notations used in this paper.}
\label{tab:notations}
\end{table*}

%% file: t-basic.tex
We begin by reminding the definition of AsIC (i.e. Asynchronous Independent Cascades Model) according to \cite{asic}, which is an extension of the IC model so the diffusion can unfold in continuous-time. It models the diffusion of information through a directed network $G=(V,E)$, where $V$ is the set of all the nodes and $E (\subset V \times V)$ is the set of all the links. For each link $(v_x,v_y)$, two real values are fixed in advance: $\rho_{v_x,v_y}$, with $0<\rho_{v_x,v_y}<1$, and $r_{v_x,v_y}$, with $r_{v_x,v_y}>0$. $\rho_{v_x,v_y}$ is referred to as the diffusion probability and $r_{v_x,v_y}$ is referred to as the time-delay parameter. The diffusion process unfolds in continuous-time and, as for IC, starts from a given set of initially activated nodes $S$. Each node $v_x$ that becomes activated at time $t$ is given a single chance to activate each of its inactive neighbours $v_y$ with probability $\rho_{v_x,v_y}$ at time $t+r_{v_x,v_y}$. The stopping condition of the process is the same as for IC, i.e. when no more activations are possible.

\begin{figure}[ht]
\begin{center}
\includegraphics[width=1.0\linewidth ]{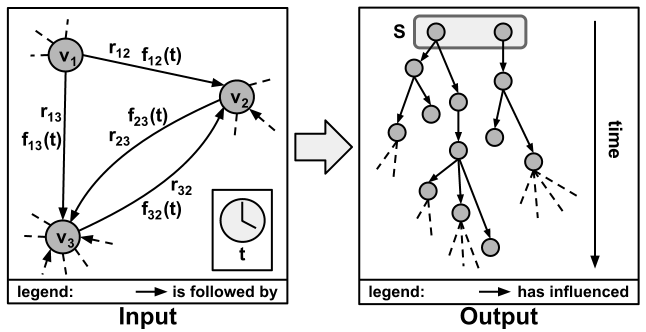}
\caption{The T-BaSIC Model predicts the cascade of diffusion along a continuous time-axis based on the time-delay and diffusion function on each edge, and the initial active node set S.}
\label{fig:tbasic}
\end{center}
\end{figure}

To enable a better capture of the dynamics underlying the diffusion process in social networks, we propose another manner in considering the parameters of the diffusion models, incorporated into the \textit{T-BAsIC} model. 
Thus, in the T-BAsIC model, a real value $r_{v_x,v_y}$ is fixed in advance and a real function $f_{v_x,v_y}(t)$ is defined for each link $(v_x,v_y)$, with $0<f_{v_x,v_y}(t)<1$. Unlike other models, the diffusion probability is not fixed in advance, but is a time dependent function $f_{v_x,v_y}(t)$ referred to as the \textit{diffusion function}. Thus, the propagation process unfolds in the same way as AsIC, but the algorithm simulates the course of days by keeping a clock and   $\rho_{v_x,v_y}$ is computed on-demand, according to $f_{v_x,v_y}(t)$. We use the model to produce the time-series that represent the evolution of the volume of tweets generated by the diffusion of a topic introduced in a given social network by a certain subset $S$ of users. Figure \ref{fig:tbasic} illustrates this principle and shows the input and output of T-BaSIC.

%% file: features.tex
Our model computes a diffusion probability relying on three dimensions: social, semantic, and time. We denote $p_{v_xv_y}(i,t)$ = $f_{v_x,v_y}(t)$ the diffusion probability of an information $i$ associated to a topic $c_i$ at time $t$ between users $v_x \in V$ (sender) and $v_y \in V$ (receiver). 
The attributes we derive from these dimensions are either numerical values varying between 0 and 1 or boolean values. Their calculation is based on the past activity of the user(s) for a given time period. Here we give the metrics formulations for a period of one month.

\textbf{Social dimension:} This dimension intends to quantify the social interactions occurring between users. It is based on metrics that mainly rely on topological properties of the \textit{active} part of the Twitter social network. This choice is motivated by the predictive power of these links in the diffusion process, as Yang and Counts stated in \cite{microsoft}. These five metrics concern whether a user or a pair of users and are described below.

\begin{itemize}
	\item \textit{Activity (I):} an activity index expresses users' volume of tweets they produce. The activity is computed as the average amount of tweets emitted per hour bounded by 1. For a user $u$, the formula is as follows:
	\begin{equation}
			\text{I}(v) = \left\{
    	\begin{array}{ll}
        	\frac{|M_{v}|}{\epsilon} & \mbox{if } |M_{v}|<\epsilon \\
        	1 & \mbox{Otherwise}
    	\end{array}
	\right.
	\end{equation}
	
	with $\epsilon = 30.4\times24$ to obtain the hourly frequency.

	\item \textit{Social homogeneity (H): } a social homogeneity index for $v_x \in V$ and $v_y \in V$ reflects the overlap of the sets of users they interact with. It is computed with the Jaccard similarity index that we defined as the size of the intersection of the sets divided by the size of their union.
	\begin{equation}
	 \text{H}(v_x, v_y) = \frac{|{\cal{M}}_{v_i} \cap {\cal{M}}_{v_j}|}{|{\cal{M}}_{v_i} \cup {\cal{M}}_{v_j}|}
	\end{equation}
			
	\item The ratio of directed tweets for each user ($dTR$) provides an idea about the role she plays in the spread of information. A user with an important ratio of directed tweets tends to play an \textit{active role} whereas a user with a low ratio can be seen as a more \textit{passive actor}. It should be noted that our definition of directed tweets includes retweets. This ratio is computed as follows:
	\begin{equation}
	\text{dTR}(v) = \left\{
    	\begin{array}{ll}
        	\frac{|{\cal{D}}_{v}|}{|M_v|} & \mbox{if } |M_v|>0 \\
        	0 & \mbox{Otherwise}
    	\end{array}
	\right.
	\end{equation}
		
	\item A boolean value for each user regarding the mentioning behaviour to capture the existence of an active interaction in the past. This feature can be somehow regarded as a ``friendship'' indicator in the case where both users have a positive value. This constitutes a different definition of friendship than the one given by Huberman et al.~\cite{microscope}, where a user is friend with users she mentioned at least twice.
	\begin{equation}
	\text{hM}(v_x, v_y) = \left\{
    	\begin{array}{ll}
        	1 & \mbox{if } v_y \in {\cal{M}}_{v_x} \\
        	0 & \mbox{Otherwise}
    	\end{array}
	\right.
	\end{equation}
	
	\item The mention rate ($mR$)~\cite{microsoft} of each user represents the volume of directed tweets she receives. Thus, the higher the value is, the higher the node centrality degree on the active part of the network is. All in all, this feature expresses the popularity of the user and the amount of information she is exposed to.
	\begin{equation}
	\text{mR}(v) = \left\{
    	\begin{array}{ll}
        	\frac{|t{\cal{M}}^v|}{\mu} & \mbox{if } |t{\cal{M}}^v|< \mu \\
        	1 & \mbox{Otherwise}
    	\end{array}
	\right.
	\end{equation}	
	
	Based on our empirical observation of the distribution of the mention rates we have chosen $\mu=200$.
	
\end{itemize}

\textbf{Semantic/Topical dimension}: In addition to the social features that exploit the structure of the network, we consider the exchanged content to refine our perception of users' behaviour. The proposed metric applies to a user and a topic and it states in a binary way if the user has already tweeted about the given topic.
	\begin{equation}
	\text{hK}(v,i) = \left\{
    	\begin{array}{ll}
        	1 & \mbox{if }c_{i}^{1} \in K_v \\
        	0 & \mbox{Otherwise}
    	\end{array}
	\right.
	\end{equation}

\textbf{Temporal dimension: } Finally, we consider the temporal dimension so that we can capture the fluctuation of users attention through time. The varying attention of the individuals is an important characteristic of online social networks that is strongly connected to the day/night cycle and working hours. To represent how the attention of a user evolves during a day, we define a receptivity function  $A(u,t)$. We model it in a non-parametric way and thus partition a day into 6 bins of 4 hours each, in order to obtain a significant and smooth representation even for less active users. We define the receptivity level of a user at the time of the day $t$ as the percentage of all the tweets she produced in the 4 hours interval $[t_x;t_y]$, where $t_x<t<t_y$. The function is stored in a 6-dimensional non-negative vector noted $V$, with $\sum_{t=0}^{5} V^{t} = 1$.
\begin{equation}
 \text{A}(v,t)=V_{v}^{t'} \text{, where } t' = \lfloor\frac{t}{4}\rfloor
\end{equation}

For instance, if a user posted 60 messages through the month, 50 of which between 4 and 8 pm and the rest between 8 and 12 am, her receptivity function would be $V_u = \{0,0,0.167,0,0.833,0\}$.

\begin{figure}[ht]
\begin{center}
\includegraphics[width=0.92\linewidth ]{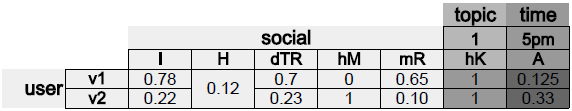}
\caption{A pair of user ($v_x$,$v_y$) is described by 13 features w.r.t a topic $c_i$ and a time of the day $t$. The figure above shows how the 3 dimensions of the feature space are connected.}
\label{fig:features}
\end{center}
\end{figure}

\textbf{Global feature space: } Overall, the metrics we just detailed form a 3-dimensional feature space, with 13 values describing each set ($v_x \in V$, $v_y \neq v_x \in V$, $c_i \in \cal{C}$, $t$). Figure \ref{fig:features} illustrates a possible instantiation of that vector.

Once the features space is constructed, the parameters of the model can be learned and estimated. This is performed through machine learning techniques and is detailed in the following section. 

%% file: diffusion_function.tex
\begin{table*}
\centering
\begin{tabular}{|c|c|c|c|c|c|}
\hline \textbf{{\small Social}} & \textbf{{\small \# of users}} & \textbf{{\small \# of tweets}} & \textbf{{\small\# of following edges}} & \textbf{{\small active network density}} & \textbf{{\small passive network density}}\\
 \textbf{{\small network}} & & \textbf{{\small (November)}} & & \textbf{{\small (November)}} & \\
\hline 1 & 24,571 & 303,564 & 1,928,999 & $5.15\e{-6}$ & $6.39\e{-3}$ \\ 
\hline 2 & 44,410 & 469,775 & 4,398,953 & $4.23\e{-6}$ & $5.13\e{-3}$ \\ 
\hline 3 & 11,614 & 169,689 & 308,849 & $7.30\e{-6}$ & $4.58\e{-3}$ \\ 
\hline 4 & 29,625 & 226,753 & 2,507,768 & $2.79\e{-6}$ & $5.71\e{-3}$ \\ 
\hline 
\end{tabular} 
\caption{Properties of the four experimental social networks.}
\label{tab:social_networks}
\end{table*}

First, we build a sample of data comprising 4 experimental social networks (sub-graphs). To build them, we first choose 4 Twitter users at random among the millions contained in the data and then by selecting all users distant from at most 2 hops according to the ``following links''. Each social network presents particular characteristics in terms of level of activity, density of the passive and active parts, and global size. See Table \ref{tab:social_networks} for details. In each network, we capture the diffusion of several topics during December 2009 and build the spreading cascades with the method we described earlier in Section \ref{sec:data}. We describe each instance of ``diffusion'' and ``non-diffusion'' by the 13 features related to the two concerned users, according to their activity during November 2009. Table \ref{tab:mean} provides the mean and standard deviation of the numerical features of that learning dataset (balanced binary dataset of 20,000 instances). 

\begin{table}[ht]
\centering
\begin{tabular}{|p{2cm}|p{1.5cm}|p{3cm}|}
\hline \textbf{Feature} & \textbf{Mean} & \textbf{Standard} \\ 
 &  & \textbf{deviation} \\ 
\hline I(src) & 0.148 & 0.185 \\ 
\hline I(dst) & 0.104 & 0.143 \\ 
\hline mR(src) & 0.163 & 0.258 \\ 
\hline mR(dst) & 0.22 & 0.324 \\ 
\hline dTR(src) & 0.488 & 0.242 \\ 
\hline dTR(dst) & 0.47 & 0.276 \\ 
\hline A(src,t) & 0.306 & 0.178 \\ 
\hline A(dst,t) & 0.247 & 0.192 \\ 
\hline H(src,dst) & 0.004 & 0.02 \\ 
\hline 
\end{tabular}
\caption{Mean and standard deviation of the numerical features of the learning dataset.}
\label{tab:mean}
\end{table}

We train several classification algorithms on the supervised task $P(Y|F)$, with $Y$=\{diffusion,non-diffusion\} and $F$ the 13-dimensional feature vector. Results obtained by a C4.5 regression tree, linear and multilayer (1 hidden layer of 14 nodes) Perceptrons and the Bayesian logistic regression (BLR) are shown on Table \ref{tab:classifiers}. All classifiers perform equally, apart from C4.5 that has a slightly better precision rate. Because the regression tree is more vulnerable to over-fitting, and linear and multilayer Perceptrons give similar results, we use the Bayesian logistic regression to define the \textit{diffusion function}.

\begin{table}[ht]
\centering
\begin{tabular}{|p{4cm}|p{3cm}|}
\hline{\bf {\small Classifier}} & {\bf
{\small Correctly classified}} \\
& {\bf {\small instances}} \\
\hline {\small C4.5}  & {\small 91\%} \\
\hline {\small Linear Perceptron} & {\small 85\%} \\
\hline {\small Multilayer Perceptron} & {\small 86\%} \\
\hline {\small Bayesian logistic regression} & {\small 85\%} \\
\hline
\end{tabular}
\caption{Classifiers performances on a 5 folds cross-validation.}
\label{tab:classifiers}
\end{table}

The BLR assumes a parametric form for the distribution $P(Y|F)$. The parametric model used by the logistic regression is as follows (as defined in~\cite{mitchell}):
$$
 P(Y=\text{diffusion}|F) = \frac{1}{1+\exp(w_0+\sum_{a=1}^{13}w_aF_a)} 
$$
$$
	P(Y=\text{non-diffusion}|F) = \frac{\exp(w_0+\sum_{a=1}^{13}w_aF_a)}{1+\exp(w_0+\sum_{a=1}^{13}w_aF_a)}
$$

In more details, the Bayesian logistic regression has a precision rate of 79\% based on the attributes belonging to the social dimension and obtains a gain of 7\% with the addition of both temporal and semantic dimensions, leading to a precision rate 85\% with the full feature space. Figure \ref{fig:BLR} illustrates the absolute normalized values of the weights (i.e. $|w_a/\max (w_a)|$) that the logistic regression accords to each feature. The ``social homogeneity'' has the highest coefficient, because it has a mean of 0.004, which is much lower than the other features. One can also see that the most significant properties of the receiving node are the level of activity and the connection to the topic. Concerning the sending node, the mention rate is the most relevant property.

\pgfplotsset{width=7.5cm,height=4.5cm}
\begin{figure}[h]
\begin{center}
\begin{tikzpicture}
\begin{axis}[ybar,
xtick \empty,
extra x ticks={0,1,2,3,4,5,6,7,8,9,10,11,12},
extra x tick labels={I(src),I(dst),hK(src i),hK(dst i),hM(src dst),hM(dst src),mR(src),mR(dst),dTR(src),dTR(dst),A(src t),A(dst t),H(src dst)},
extra x tick style={
tick label style={rotate=90,anchor=east}},
ylabel=normalized coefficient,
bar width=5.5pt]
\addplot [black,fill=black!30!black,opacity=0.5] table 
[x=attribute, y=BLR] {BLR_LP.dat};
\end{axis}
\end{tikzpicture}
\caption{Visualization of the normalized weights computed by the Bayesian logistic regression.}
\label{fig:BLR}
\end{center}
\end{figure}
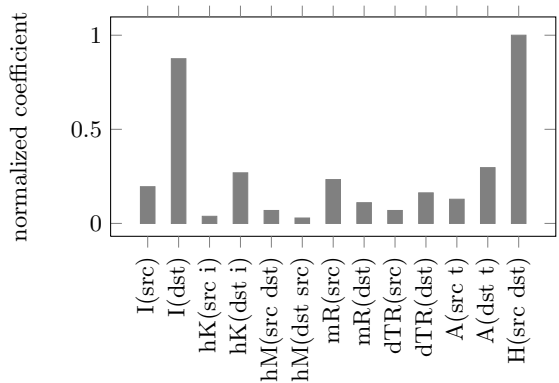

%% file: time-delay.tex
For each instance of the ``diffusion'' class from the dataset, we know at what time the two users adopted the topic (i.e. tweeted about it), so we are able to compute the real diffusion delay. As we have just seen, the activity index of the receiving user is a critical parameter and we base the approximation of the time-delay on it, with the following formula: $r_{v_x, v_y} = (1-I(v_y)) \times \sigma$. Therefore, for each instance of diffusion we can compare real and estimated diffusion delay. In order to determine the optimal value of $\sigma$, we define two vectors, (i) the vector of real diffusion delays and (ii) the vector of estimated diffusion delays. Then we define $E(\sigma)$ the Euclidean distance between these two vectors depending of the value of $\sigma$. We find that $E(\sigma)$ is minimal for $\sigma = 10$ and so the formula used to estimate the diffusion delay becomes $r_{v_x, v_y} = (1-I(v_y)) \times 10$. It means that the maximum diffusion delay in our modeling is of 10 hours. This aligns well with observations made in previous studies \cite{galuba,lim}, that reveal that diffusion events occur across a time-frame of at most 8 to 12 hours.

%% file: prediction_engine.tex
Having established how to estimate T-BaSIC parameters, we use it as a prediction engine. The required input is: (1) a topic, as defined in Section \ref{sec:data}; (2) a social network described by: (i) the set of users $V$ and their 3-dimensional description, (ii) the topology of their interconnection based on the following links; and (3) a subset of users $S \subset V$ that inject the topic in the network and thus initiate the diffusion.
Given this input, the algorithm unfolds and manages a clock, which is used to reproduce the course of day and the variations of users receptivity. In output, the engine generates time-series representing the volume of tweets induced by the spread of the topic inside the network.

%% file: generalization.tex
T-BaSIC is a generic model, but the estimation of its parameters depends on the social platform one wants to adapt it to. The approach we have presented can be applied to any social network based on the explicit declaration of social links and that permits its users to publish both directed and undirected messages. The coefficients of the diffusion function and the time-delay can be then adjusted to the data during the learning step.

%% file: experiments.tex
We evaluate the efficiency of our approach and modeling on the task of predicting the temporal dynamics of the spread of topics selected with the method described in Section~\ref{sec:data}. We denote $P_{c_i}(t)$ the predicted daily volume of tweets for topic $c_i$ in the network and $R_{c_i}(t)$ the real daily volume of tweets (i.e. observed in the data). The different networks used for the experiments are described in Table \ref{tab:social_networks}. We choose the users constituting the starting set $S$ by selecting the first $s$ users observed in $R_{c_i}(t)$. Hereafter we present the results we obtained with the optimal value of $s$ for selected examples.

%% file: qualitative_results.tex
Figure \ref{fig:compiphone} shows the comparison of real and predicted time-series for the topic \{``iphone'',``release''\} in experimental networks \#1 and \#2. The $x$ axis represents time units in days and the $y$ axis represents the activity level with tweets volume as unit. The gray dashed curve corresponds to the real volume measured in the data while the black curve corresponds to the volume predicted by T-BaSIC. After varying the value of $s$ in the experiments, the optimal prediction is obtained using $s=8$ for network \#1 and $s=5$ for network \#2. In both cases we observe a particular wave pattern, with different phase and amplitude. One can see that the model accurately captures these variations but slightly underestimates the volume. We examine more in details the prediction made by T-BaSIC by analysing the population of users involved throughout the diffusion process. We classify them into two groups, based on definitions by Daley and Kendall \cite{DK}: (i) transmitters, i.e. users who received the information and then transmitted it to others, and (ii) stiflers, i.e. users that received the information but never transmitted it. We show on Figure \ref{fig:TS} the evolution of the density of stiflers for the prediction made on network \#1. The correlation between the volume shape and the density of stifler is clearly visible. Five days after the appearance of the information, the density of stifler is continuously rising. This is due in part to the low connectivity of these users. Indeed, they are reached by the information later in the process and have a lower potential of diffusion. This shows the relevance of the graph-based approach.

\pgfplotsset{width=4.5cm,height=4cm }
\begin{figure}[ht]
\begin{center}
\begin{tikzpicture}
\begin{axis}[name=plot1,
xlabel=Time (days),
ylabel=Volume of tweets,
title=\textbf{(a) network \#1}]
\addplot [line width = 1pt,smooth,black,opacity=0.7,dashed] coordinates {
(1,16)
(2,66)
(3,69)
(4,21)
(5,17)
(6,7)
(7,34)
(8,46)
(9,26)
};
\addplot [line width = 1pt,smooth,black,opacity=0.95] coordinates {
(1,8)
(2,48)
(3,40)
(4,14)
(5,3)
(6,0)
(7,21)
(8,29)
(9,8)
};
\end{axis}

\begin{axis}[name=plot2,at={($(plot1.east)+(0.7cm,0)$)},anchor=west,
xlabel=Time (days),
title=\textbf{(b) network \#2},legend style={
at={(-0.25,-0.45)},
anchor=north,
legend columns=-1}]
\addplot [line width = 1pt,smooth,black,opacity=0.7,dashed] coordinates {
(1,5)
(2,27)
(3,13)
(4,7)
(5,2)
(6,14)
(7,9)
(8,8)
};
\addplot [line width = 1pt,smooth,black,opacity=0.95] coordinates {
(1,5)
(2,21)
(3,11)
(4,4)
(5,1)
(6,6)
(7,0)
(8,0)
};
\legend{$R_{c_i}$,$P_{c_i}$}
\end{axis}
\end{tikzpicture}
\caption{Comparison of real and predicted time-series for the topic \{``iphone'',``release''\} in experimental networks \#1 and \#2.}
\label{fig:compiphone}
\end{center}
\end{figure}
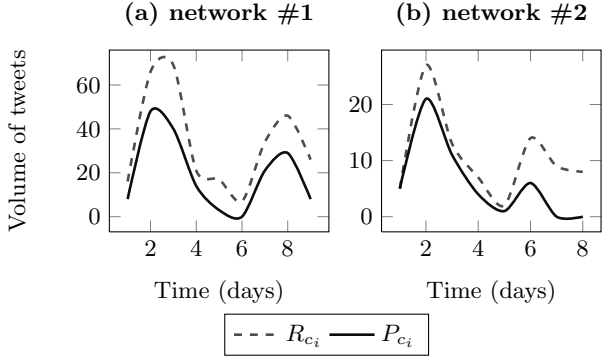
\pgfplotsset{width=5.5cm,height=4cm }
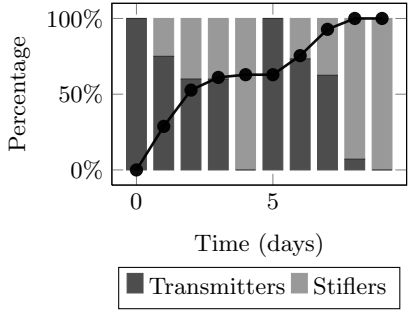
\begin{figure}[h]
\begin{center}
\begin{tikzpicture}
\begin{axis}[ybar stacked, legend style={area legend,
at={(0.5,-0.45)},
anchor=north,
legend columns=-1},
xlabel=Time (days),
ylabel=Percentage,
ytick \empty,
extra y ticks={0,50,100},
extra y tick labels={0\% ,50\% ,100\% },
bar width=7.5pt]
\addplot [black,fill=black!30!black,opacity=0.7] table 
[x=Time, y=Tp] {ITS-prop.dat};
\addlegendentry{Transmitters}
\addplot [black,fill=black!30!black,opacity=0.4] table 
[x=Time,
y=Sp] {ITS-prop.dat};
\addlegendentry{Stiflers}
\end{axis}
\begin{axis} [ytick \empty,xtick \empty]
\addplot [line width = 1pt,black,opacity=0.95, mark=*] coordinates {
(0,8)
(1,56)
(2,96)
(3,110)
(4,113)
(5,113)
(6,134)
(7,163)
(8,175)
(9,175)
};
\end{axis}
\end{tikzpicture}
\caption{Density of transmitters and stiflers. The curve represents the cumulation of the number of users that adopted the topic.}
\label{fig:TS}
\end{center}
\end{figure}
\pgfplotsset{width=4.5cm,height=4cm }
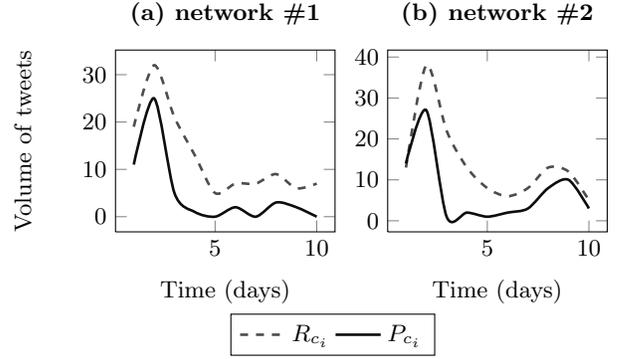
\begin{figure}[h]
\begin{center}
\begin{tikzpicture}
\begin{axis}[name=plot1,
xlabel=Time (days),
ylabel= Volume of tweets,
title=\textbf{(a) network \#1}]
\addplot [line width = 1pt,smooth,black,opacity=0.7,dashed] coordinates {
(1,19)
(2,32)
(3,21)
(4,13)
(5,5)
(6,7)
(7,7)
(8,9)
(9,6)
(10,7)
};
\addplot [line width = 1pt,smooth,black,opacity=0.95] coordinates {
(1,11)
(2,25)
(3,5)
(4,1)
(5,0)
(6,2)
(7,0)
(8,3)
(9,2)
(10,0)
};
\end{axis}

\begin{axis}[name=plot2,at={($(plot1.east)+(0.7cm,0)$)},anchor=west,
xlabel=Time (days),
title=\textbf{(b) network \#2}, legend style={
at={(-0.25,-0.45)},
anchor=north,
legend columns=-1}]
\addplot [line width = 1pt,smooth,black,opacity=0.7,dashed] coordinates {
(1,13)
(2,38)
(3,22)
(4,13)
(5,8)
(6,6)
(7,8)
(8,13)
(9,12)
(10,5)
};
\addplot [line width = 1pt,smooth,black,opacity=0.95] coordinates {
(1,14)
(2,27)
(3,1)
(4,2)
(5,1)
(6,2)
(7,3)
(8,8)
(9,10)
(10,3)
};
\legend{$R_{c_i}$,$P_{c_i}$}
\end{axis}
\end{tikzpicture}
\caption{Comparison of real and predicted time-series for the topic \{``google'',``buy''\} in experimental networks \#1 and \#2.}
\label{fig:compgoogle}
\end{center}
\end{figure}

In order to allow comparison, we now show the results obtained for another topic, \{``google'',``buy''\} in the same two networks. After varying the value of $s$ in the experiments, the optimal prediction is obtained using $s=11$ for network \#1 and $s=14$ for network \#2. Again, we can see a wave-pattern, but this time it is less strong, which reveals people are less interested by this topic. This highlights the importance of taking into account the topical dimension into the computation of the diffusion probabilities. Also, once again, the predicted volume is inferior to the real volume. 

We obtained similar results with all the selected topics and predictions on the four experimental social networks, with the size of $S$ varying between 5 and 20. It means that, starting from a few informed people we are able to predict when peaks of attention will occur and in which proportion. This result is consistent with the ``two-step flow of communication'' theory introduced by Katz and Lazarsfeld \cite{two_step}, that hypothesizes information flows from media to few ``opinion leaders'' who then spread it to the ``mass'' via social networks. However we found that our modeling always underestimates the global volume. This can be explained by the fact that we consider that individuals, apart from those constituting the starting set $S$, get information exclusively via their social network, as explicitly stated in the problem formulation referred to as the ``\textit{closed environment}''. But in reality, information is injected into Twitter throughout the diffusion process and not only at the beginning.

%% file: quantitative_results.tex
\pgfplotsset{width=3.5cm,height=2.5cm }
\begin{table*}[ht]
\centering
\begin{tabular}{|l|p{2cm}|p{2cm}|p{2cm}|p{2cm}|r|}
\hline  & Shape 1 & Shape 2 & Shape 3 & Shape 4 & ALL \\ 
 
 & \begin{tikzpicture}
\begin{axis} [ytick \empty,xtick \empty]
\addplot [line width = 0.5pt,smooth,black,opacity=0.7] coordinates {
(1,16)
(2,66)
(3,69)
(4,21)
(5,17)
(6,7)
(7,34)
(8,46)
(9,26)
};
\end{axis}
\end{tikzpicture}
 & \begin{tikzpicture}
\begin{axis} [ytick \empty,xtick \empty]
\addplot [line width = 0.5pt,smooth,black,opacity=0.7] coordinates {
(1,5)
(2,27)
(3,13)
(4,7)
(5,2)
(6,14)
(7,9)
(8,8)
};
\end{axis}
\end{tikzpicture}
 & 
\begin{tikzpicture}
\begin{axis}[ytick \empty,xtick \empty]
\addplot [line width = 0.5pt,smooth,black,opacity=0.7] coordinates {
(1,19)
(2,32)
(3,21)
(4,13)
(5,5)
(6,7)
(7,7)
(8,9)
(9,6)
(10,7)
};
\end{axis}
\end{tikzpicture} 
&
\begin{tikzpicture}
\begin{axis}[ytick \empty,xtick \empty]
\addplot [line width = 0.5pt,smooth,black,opacity=0.7] coordinates {
(1,13)
(2,38)
(3,22)
(4,13)
(5,8)
(6,6)
(7,8)
(8,13)
(9,12)
(10,5)
};
\end{axis}
\end{tikzpicture} 
 & \\
\hline reduction on dynamics & 25.19\% & 39.23\% & 29.21\% & 3.22\% & 24.21\% \\ 
\hline reduction on volume & 42.89\% & 47.70\% & 34.49\% & 40.07\% & 41.29\% \\ 
\hline overall gain & 34.04\% & 43.46\% & 31.85\% & 21.65\% & 32.75\% \\ 
\hline 
\end{tabular} 
\caption{Reduction in prediction error on volume and dynamics over the 1-time lag predictor for four shapes of volume over time.}
\label{tab:quantitative}
\end{table*}

We now quantitatively asses the efficiency of our modeling by computing the reduction in prediction error over the 1-time lag predictor \cite{lim}, according to two aspects: (i) volume and (ii) dynamics. The 1-time lag predictor, introduced by Yang and Leskovec, is a simple predictor that gives $P_{c_i}^{'}(t)$, such as $P_{c_i}^{'}(t) = R_{c_i}(t-1)$. We compute the relative error on volume estimation with the formula below:

$$
volume\_ error_{c_i} = \frac{\sqrt{\sum_{t}(P_{c_i}(t)-R_{c_i}(t))^2 }}{\sqrt{\sum_{t}R_{c_i}(t)^2 }}
$$

We then compute the relative error on dynamics according to this formula:  

$$
dynamics\_error_{c_i}= \frac{\sqrt{\sum_{t}(d(R_{c_i}(t))-d(P_{c_i}(t)))^2}}{\sqrt{\sum_{t}R_{c_i}(t)^2 }}
$$
where $d$ is the derivative for each point of the time-series that we compute in this way: $$d(R_{c_i}(t)) = \frac{R_{c_i}(t+1)-R_{c_i}(t-1)}{2}$$

We report the reduction in prediction error on volume and dynamics of our approach over the 1-time lag predictor in Table \ref{tab:quantitative} for four particular shapes of volume over time. These shapes correspond to the examples we just detailed.
Overall, and as we can see it, the results are satisfactory, translated by the overall gain measure.

%% file: conclusion.tex
In this paper, we propose the T-BAsIC model and its application to data issued from Twitter. To achieve this, we determined with a preliminary study a set of pertinent features that ensure a generic model for representing information diffusion whose parameters are estimated with the considered data themselves. Indeed, this model allows to predict information diffusion taking into account both social, semantic, and temporal dimensions. More precisely, this model is derived from the AsIC \cite{asic} theoretical model and relies on time-dependent parameters. We infer the diffusion probabilities between nodes of the network with a machine learning technique, i.e. Bayesian logistic regression. We performed a set of experiments for different topics. The experimental results show mainly that the model predicts well the dynamic of the diffusion (our initial objective). 
The prediction of the volume is slightly underestimated due to our initial assumption of considering a ``\textit{closed environment}''. This ignores the impact of external information sources on the networks, which may explain the gap in the predicted volume and the observed volume. Still, our results support the ``two-step theory'' that hypothesizes that only a few "opinion leaders" relay information from media to the ``mass population'' via social networks and show that it also apply to \textit{online} social networks.

The perspectives opened by this work are numerous. Among them, we determined the four main issues we want to investigate. First, since the T-BAsIC model parameters are not fixed in advance, it should allow us to take into account the evolution, over time, of the environment for the estimation of diffusion probabilities. Thus it could help us to consider the phenomenon of ``complex contagion'' as introduced by \cite{complexcontagion} (i.e. repeated exposures to a topic have a positive impact on the probability that the user adopts it). Concerning the genericity of the model, another issue consists in applying our approach on other social data from other platforms to study both common points and specificities of the information diffusion process according to the platform. A third one consists in enriching the semantic dimension. The use of text mining techniques could be useful for this challenge, taking into account that depending on the social platform, there are some specificities that must be taken into account. Finally, T-BAsIC is based on the AsIC theoretical model, that means it is built from the point of view of the diffuser node. It could be interesting to envision the dual approach considering a T-BAsLT model based on the AsLT \cite{asic} approach focusing on the point of view of receiving node. The comparison of the two approaches could be of interest and may bring interesting insights.